\documentclass[aps,pre,superscriptaddress,twocolumn,longbibliography]{revtex4}
\usepackage{graphicx,epstopdf,url}
\usepackage[unicode,colorlinks,linkcolor=blue,citecolor=blue,urlcolor=blue]{hyperref}
\usepackage{ulem}

\begin{document}
\title{
Stability of Planar Slits in Multilayer Graphite Crystals
}
\author{A. V. Savin}
\email[]{asavin@chph.ras.ru}
\affiliation{
Semenov Federal Research Center for Chemical Physics of the Russian Academy of Sciences.
(FRC CP RAS), Moscow 119991, Kosygina st., 4}
\affiliation{
Plekhanov Russian University of Economics,\\
Moscow 117997, Stremyanny lane, 36
}
\author{A. P. Klinov}
\email[]{aklinov@chph.ras.ru}
\affiliation{
Semenov Federal Research Center for Chemical Physics of the Russian Academy of Sciences.
(FRC CP RAS), Moscow 119991, Kosygina st., 4}

\date{\today}

\begin{abstract}
Using a two-dimensional coarse-grained chain model, planar slits in multilayer graphite crystals are simulated.
It is shown that when covering a linear cavity on the flat surface of a graphite crystal with a multilayer graphene sheet, an open (unfilled slit) can form only if the cavity width does not exceed a critical value $L_o$ (for width $L>L_o$, only a closed state of the slit is formed, with the cavity space filled by the covering sheet).
The critical width of the open slit $L_o$ increases monotonically with the number of layers $K$ in the covering sheet.
For a single-layer cavity, there is a finite critical value of its width $L_o<3$~nm, while
for two- and three-layer cavities, the maximum width of the open slit increases infinitely with increasing $K$ as a power function $K^\alpha$ with exponent $0<\alpha<1$.
Inside the crystal, two- and three-layer slits can have stable open states at any width.
For a slit with width $L>7.6$~nm, a stationary closed state is also possible, in which its lower and upper surfaces adhere to each other.
Simulation of thermal oscillations showed that open states of two-layer slits with width $L<15$~nm
are always stable against thermal oscillations, while wider slits at $T>400$~K transition from the open to the closed state. 
Open states of three-layer slits are always stable against thermal oscillations.

\end{abstract}

\maketitle

\section{Introduction \label{sec1}}
Nanoscale pores and capillaries are actively studied due to their importance for understanding many natural phenomena and their potential applications. For instance, fluid that is transported through them can possess new properties, impossible on larger scales \cite{Schoch2008}.
Nanopores are used to study the biophysics and chemistry of single molecules \cite{Howorka2009}.
A method for creating smooth capillaries with precisely controlled dimensions was proposed, based on van der Waals assembly \cite{Geim2013} from two atomically flat sheets separated by spacers made of two-dimensional crystals \cite{Novoselov2005}.
Graphene and its sheets with a precisely controlled number of layers were used as the two-dimensional materials \cite{Radha2016}. Such assembly allows creating structures that can be viewed as flat empty slits a few atoms high within a graphite crystal \cite{Geim2021}.
Such slits have been used to study water transport through channels with heights ranging from one \cite{Gopinadhan2019} to several tens of atomic layers \cite{Radha2016,Keerthi2021}.
A programmable nanofluidic switch based on such channels has been proposed \cite{Ronceray2024}.

Internal cavities can also be used as nanoscale pressure sensors \cite{Sorkin2011,Smith2013,Sanaeepur2017,Ahn2017}.
Planar slits in layered structures can serve as efficient optical waveguides \cite{Ling2022}. 
Other uses for two-dimensional slits are given in the review \cite{Ma2023}.

When creating nanoslits in graphene and other layered materials, there is the problem of the stability of the obtained pores to environmental influences.
For example, in one of the early works, nanochannels one atomic layer high were not stable and collapsed in the experiment \cite{Radha2016}.
In fact, only nanochannels with heights of two, three, or more layers could be obtained.
As shown in a later work, the inability to obtain nanochannels one layer high was related to insufficiently careful processing of the edges of the graphite nanocrystal, which caused channel closure \cite{Gopinadhan2019}.
Authors of the work \cite{Ronceray2024} proposed a phenomenological model to describe the stability of nanoslits.
In this model, pore stability is determined by the balance of van der Waals interactions between the upper and lower pore surfaces, the deformation energy of the pore upon collapse, and the capillary pressure caused by nanocondensation of the solvent inside the channel.
Although this model yields results that agree reasonably well with experiment, it pays little attention to the case of a small number of layers in the upper part of the pore (a continuum approximation is applied), and also does not account for the influence of layer sliding and thermal fluctuations.
In this work, by using a coarse-grained and all-atom model of graphene pores, the influence of these factors on the stability of nanopores will be considered in more detail.

Graphene sheets (nanoribbons) can easily bend and slide relative to each other, allowing them to conform to substrate irregularities \cite{Han2020}. 
Such high mobility of layers prevents the formation of large planar cavities (slits) with a height of several layers, because due to displacement and bending, the upper and lower layers of the cavity can approach each other, filling its space \cite{Savin2023pe}.
In this work, using a two-dimensional coarse-grained chain model of a multilayer crystal, the maximum possible cavity sizes will be determined.
It will be shown that the allowable sizes depend on the number of layers in the graphene sheets forming the cavity surface.

Part \ref{sec2} describes the two-dimensional coarse-grained chain model of a multilayer graphite crystal as a system of parallel linear chains.
Section \ref{sec3} uses this model to find stationary states of planar slits formed when covering a cavity on a flat crystal surface with a multilayer graphene sheet. 
Part \ref{sec4} models planar slits inside a graphite crystal, and part \ref{sec5} analyzes their stability against thermal oscillations.
Section \ref{sec6} performs simulations using all-atom 3D models to verify the obtained results. 
The main conclusions are presented in the Section \ref{sec7}.

\section{Chain Model of a Multilayer Graphene Sheet \label{sec2}}

\begin{figure}[tb]
\begin{center}
\includegraphics[angle=0, width=1.\linewidth]{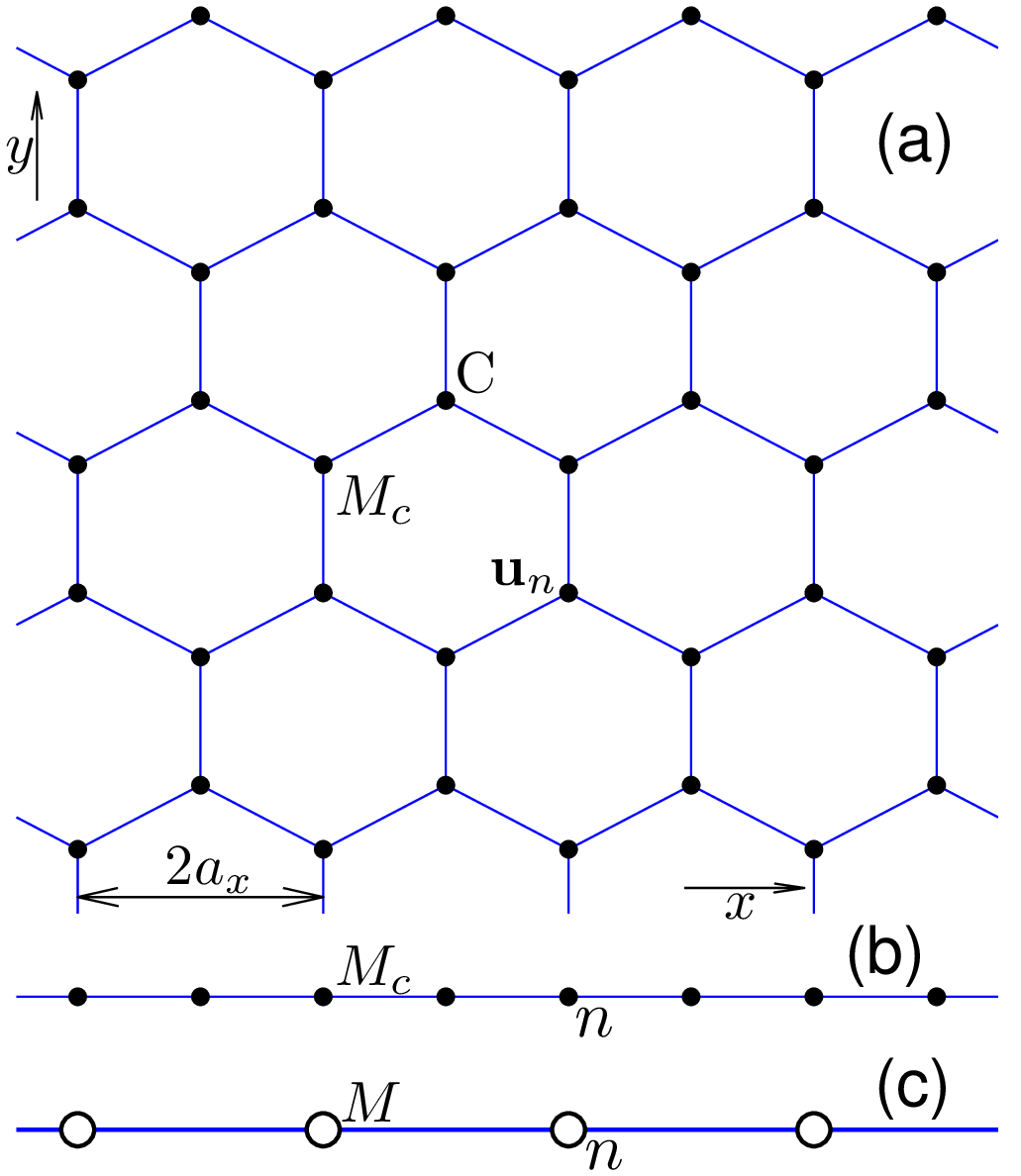}
\end{center}
\caption{\label{fg01}\protect
Schematic of the two-dimensional chain model construction for (a) a flat graphene sheet lying in the $xy$ plane (${\bf u}_n$ is the coordinate vector of the $n$-th carbon atom).
Chain model (b) with period $a=a_x$ and (c) $a=2a_x$ ($n$ is the bead index, $M_c$ is the carbon atom mass, $M=2M_c$ is the bead mass).
}
\end{figure}

A graphene sheet is an elastically isotropic material; its longitudinal and bending stiffnesses weakly depend on its chirality.
Therefore, for definiteness, let us consider deformations of the sheet in the zigzag direction -- see Fig. \ref{fg01},a.

Let the sheet lie in the $xy$ plane of three-dimensional space in its ground state. 
In the direction of the $x$-axis (the zigzag direction), the sheet is a periodic structure with period $a_x=r_c\cos(\pi/6)$, where $r_c=1.418$~\AA~ is the equilibrium length of the C--C valence bond. 
The translational unit cells of this structure are formed by atoms located along lines parallel to the $y$-axis.
If we consider motions of the graphene sheet where atoms located on the same vertical line move synchronously, then the dynamics of the sheet can be described as displacements of a linear chain of beads in the $xz$ plane -- see Fig. \ref{fg01},b.
Here, each bead of the chain describes the displacements of all atoms of the sheet located on the same vertical line (all atoms having the same coordinates $x$, $z$).
Such a two-dimensional chain model was previously used to simulate the dynamics of rolled carbon nanoribbons \cite{Savin2015prb,Savin2015ftt} and folds in a graphene sheet lying on a flat substrate \cite{Savin2019prb}.
\begin{figure}[tb]
\begin{center}
\includegraphics[angle=0, width=1.\linewidth]{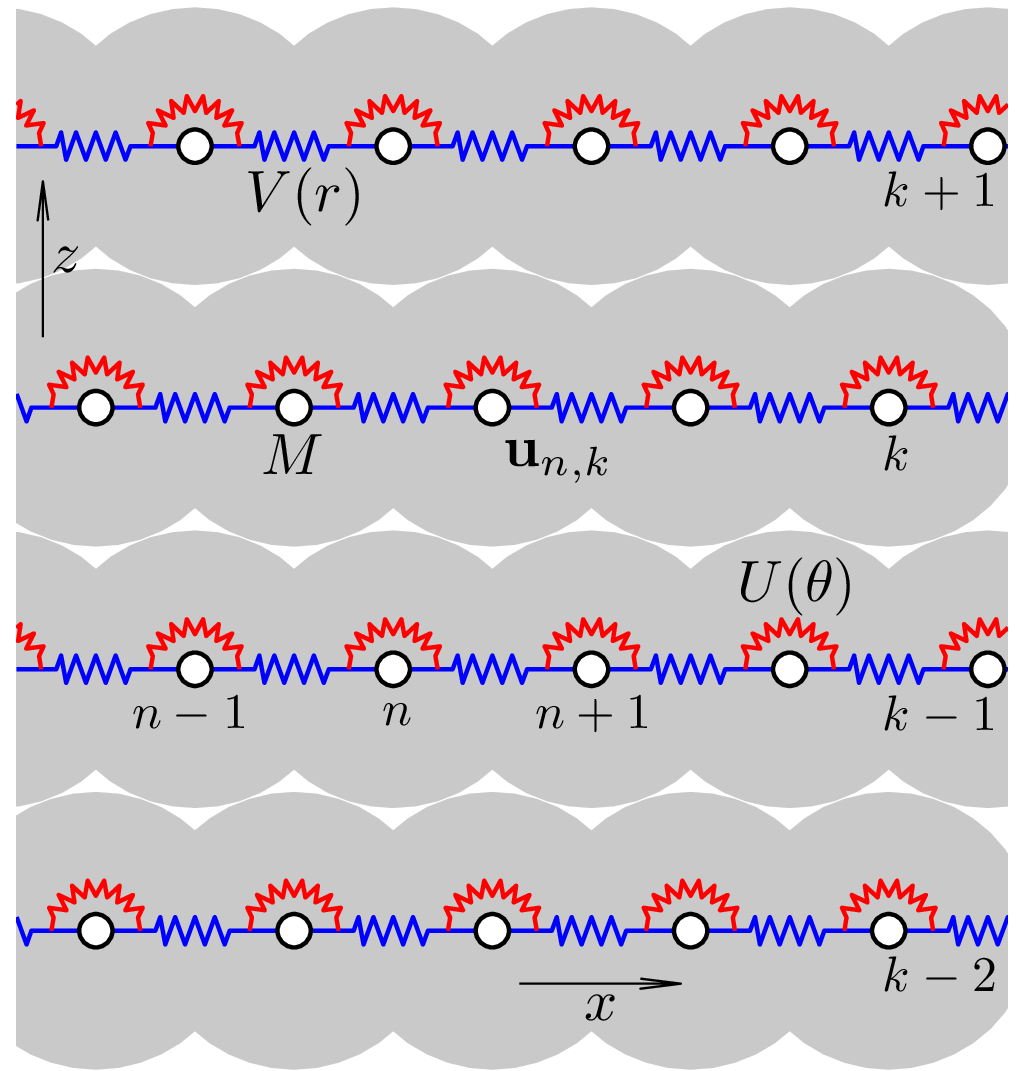}
\end{center}
\caption{\label{fg02}\protect
Chain model of a multilayer graphene sheet, $k$ is the index of chain (or sheet layer), $n$ is the bead index (vector ${\bf u}_{n,k}$ gives the coordinates of bead $n,k$), $M$ is the bead mass.
Potential $V(r)$ describes the longitudinal stiffness, and $U(\theta)$ describes the bending stiffness of the chain.
Gray circles show the van der Waals radii of the beads.
}
\end{figure}

The Hamiltonian of such a chain has the form
\begin{equation}
H=\sum_n\left[\frac12M_c(\dot{\bf u}_n,\dot{\bf u}_n)+V(r_n)+U(\theta_n)\right],
\label{f1}
\end{equation}
where $n$ denotes bead index, $M_c=12m_p$ is the mass of a bead (mass of a carbon atom), ${\bf u}_{n}=(x_{n},z_{n})$ is the vector specifying the position of the $n$-th bead, $r_{n}=|{\bf v}_{n}|$ is the distance between neighboring beads $n$ and $n+1$ (vector ${\bf v}_{n}={\bf u}_{n+1}-{\bf u}_{n}$), $\theta_{n}$ is the angle between adjacent beads (angle between vectors ${\bf v}_{n}$ and $-{\bf v}_{n-1}$).

The longitudinal stiffness of the chain is described by a harmonic potential
\begin{equation}
V(r)=\frac12K_r(r-a)^2,
\label{f2}
\end{equation}
where $a=a_x=1.228$\AA~is the chain period, $K_r=405$~N/m is the inter-bead interaction stiffness.
The bending stiffness of the chain is described by the potential
\begin{equation}
U(\theta)=\epsilon_\theta[\cos(\theta)+1],
\label{f3}
\end{equation}
where the cosine of the $n$-th angle $\cos(\theta_n)=-({\bf v}_{n-1},{\bf v}_n)/r_{n-1}r_n$, energy $\epsilon_\theta=3.5$~eV.
With these values of $K$ and $\epsilon_\theta$, the dispersion curves of the chain most accurately coincide with the dispersion curves of a flat graphene nanoribbon corresponding to its longitudinal and bending oscillations \cite{Savin2015prb}.

In the chain model, a multilayer graphene sheet corresponds to a system of parallel chains (Fig. \ref{fg02}) with the
Hamiltonian
\begin{equation}
H=\sum_{k=1}^K\sum_{n=1}^N\frac12 M_c(\dot{\bf u}_{n,k},\dot{\bf u}_{n,k})+E_1+E_2,
\label{f4}
\end{equation}
where $k$ denotes the chain index, and $n$ denotes the bead index.
The vector ${\bf u}_{n,k}=(x_{n,k},z_{n,k})$ specifies the coordinates of the $n$-th bead of the $k$-th chain ($K$ is the number of chains, $N$ is the number of beads in a chain).

The first term in the Hamiltonian (\ref{f4}) gives the kinetic energy of the molecular system, the second term describes the potential energy of the chains
\begin{equation}
E_1=\sum_{k=1}^K\sum_{n=1}^N [V(r_{n,k})+U(\theta_{n,k})],
\label{f5}
\end{equation}
where the distance between neighboring beads of the chain is $r_{n,k}=|{\bf v}_{n,k}|$ (vector  ${\bf v}_{n,k}={\bf u}_{n+1,k}-{\bf u}_{n,k}$), cosine of the valence angle $\cos\theta_{n,k}=-({\bf v}_{n-1,k},{\bf v}_{n,k})/r_{n-1,k}r_{n,k}$.
The third term
\begin{equation}
E_2=\sum_{k_1=1}^{K-1}\sum_{k_2=k_1+1}^{K}\sum_{n_1=1}^N\sum_{n_2=1}^N W(r_{n_1,k_1;n_2,k_2}),
\label{f6}
\end{equation}
describes weak non-valent (van der Waals) inter-chain interactions, $r_{n_1,k_1;n_2,k_2}=|{\bf u}_{n_2,k_2}-{\bf u}_{n_1,k_1}|$ is the distance between beads $(n_1,k_1)$ and $(n_2,k_2)$.
The interaction energy can be described with high accuracy by the (5,11) Lennard-Jones potential
\begin{equation}
W(r)=\epsilon_0\left[5(r_0/r)^{11}-11(r_0/r)^5\right]/6,
\label{f7}
\end{equation}
with equilibrium bond length $r_0=3.61$~\AA, interaction energy $\epsilon_0=0.0083$~eV \cite{Savin2019prb}.

The homogeneous state of the multilayer sheet (Fig. \ref{fg02}) is given by the coordinates $\{ {\bf u}_{n,k}^0=(x_{n,k}^0,z_{n,k}^0)\}_{n=1,k=1}^{N,~K}$, where $x_{n,k}^0=na$ for odd $k$ and $x_{n,k}^0=(n+\frac12)a$ for even $k$, $z_{n,k}^0=kh_0$, $h_0=3.352$~\AA~ is the distance between neighboring chains (layers).
For modeling an infinite sheet, it is convenient to use periodic boundary conditions, and for a sheet of finite size -- free boundary conditions.

To find the ground state, one needs to solve the problem of minimizing the potential energy of the molecular system
\begin{equation}
E=E_1+E_2\rightarrow\min : \{ {\bf u}_{n,k}\}_{n=1,k=1}^{N,~K}.
\label{f8}
\end{equation}
The minimization problem (\ref{f8}) was solved numerically by the conjugate gradient method \cite{Fletcher1964,Shanno1976}.
For the chain system corresponding to crystalline graphite, the distance between neighboring chains (layers) is $h_0=3.352$~\AA, and the interaction energy of one bead with a neighboring chain is $E_0=-52.05$~meV.
Here, to shift a chain by one period, an energy barrier of $\Delta E=0.1$~meV per bead must be overcome.
For comparison, the adhesion and pinning energies calculated within the density functional theory (DFT) approximation are $E_0=49$~meV/atom, $\Delta E=6$ meV/atom, respectively \cite{Ouyang2018nl}.
Thus, the two-dimensional chain model describes the interaction energy of graphite sheets quite accurately, but due to the very low pinning energy $\Delta E$, it allows chains to slide almost freely relative to each other.
In three-dimensional models describing the interaction energy of atoms in neighboring graphene layers using the Lennard-Jones (6,12) potential \cite{Setton1996}, pinning energy is also underestimated ($\Delta E=0.43$~meV/atom).
Exact agreement with DFT data can be obtained in models with the Kolmogorov-Crespi potential for interlayer interactions of atoms \cite{Kolmogorov2005,Maaravi2017,Ouyang2018nl}.
This potential accounts for the dependence of the interaction energy on the mutual orientation of the normals to the graphene sheet, allowing a more accurate description of the sliding energy surfaces.
\begin{table}[tb]
\caption{
Changes in the parameters of the chain model when the chain period is increased by a factor of $d$ ($\Delta E$ is the pinning energy).
\label{tab1}
}
\begin{tabular}{c|ccccccc}
\hline
\hline
~$d$~ & $a$  & $M$ & $K_0$~(N/m) & $\epsilon_\theta$ (eV) & $\epsilon_0$ (eV) & $r_0$ (\AA) & $\Delta E$ (meV) \\
 1  &  $a_x$  & $M_c$ &  405.0  & 3.500 & 0.0083 & 3.610  & 0.1 \\
 2  &  $2a_x$ & $2M_c$ &  202.5  & 1.750 & 0.0316 & 3.673 & 4.3 \\
 2.129& $da_x$ & $dM_c$ &  190.2  & 1.644 & 0.0352 & 3.694 & 6.0\\
\hline
\hline
\end{tabular}
\end{table}

In the chain model, the pinning energy can be increased by increasing the chain period $a$ (by using a more coarse-grained chain).
Let us introduce the discreteness parameter $d=a/a_x$. 
In order to preserve the linear density, longitudinal and bending stiffness when increasing the chain period by a factor of $d$, the mass of a chain link must be increased by a factor of $d$, and the parameters $K$ and $\epsilon_\theta$ must be decreased by a factor of $d$. 
The parameters of the potential (\ref{f7}) must be changed to preserve the ground state, the distance between neighboring chains $h_0$ and their interaction energy $E_0$ per chain length $a_x$ in the ground state.
Specific values of the changed parameters are given in Table \ref{tab1}.
Of course, when increasing the chain discreteness, the positions of beads no more correspond to positions of carbon atoms of the graphene sheet, but in the continuum approximation, the deformations of the chain will continue to correspond to the deformations of the sheet.
\begin{figure*}[tb]
\begin{center}
\includegraphics[angle=0, width=0.75\linewidth]{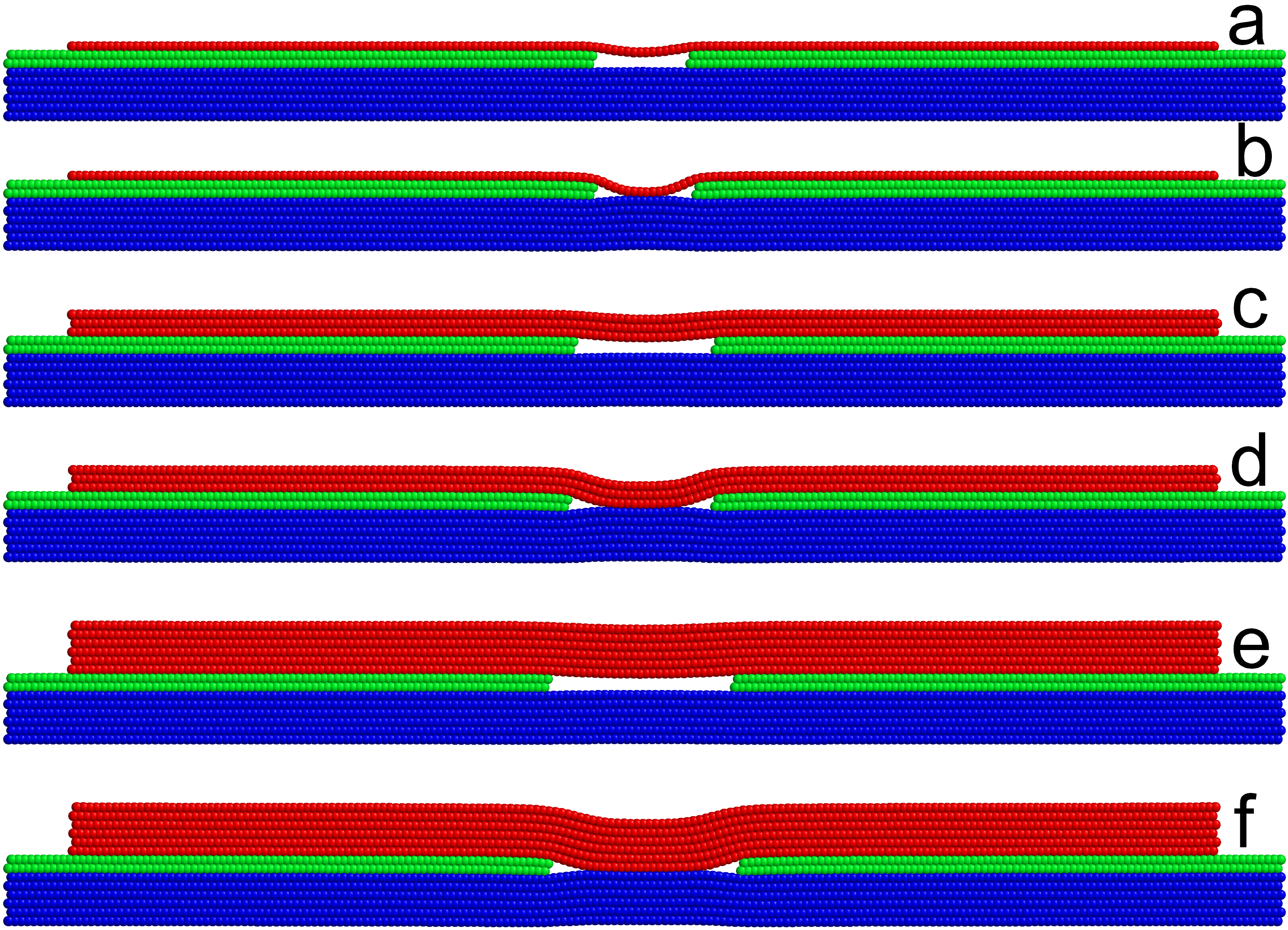}
\end{center}
\caption{\label{fg03}\protect
Stationary state of a two-layer slit ($K_g=2$) of width $L_x=(N_g+1)a$, covered by a $K$-layer graphene sheet for (a) $N_g=15$, $K=1$; (b) $N_g=16$, $K=1$, (c) $N_g=22$, $K=3$, (d) $N_g=23$, $K=3$, (e) $N_g=29$, $K=6$, (f) $N_g=30$, $K=6$.
The upper graphene sheet is shown in red, the substrate sheets participating in the slit formation are shown in green, the remaining substrate layers are shown in blue.
The chain model with discretization parameter $d=2$ (chain period $a=2a_x=2.456$\AA) is used.
Only the top 8 layers of the substrate out of $K_s=50$ layers are shown (number of beads in the chain $N_s=200$, $N=180$).
}
\end{figure*}
\begin{figure}[tb]
\begin{center}
\includegraphics[angle=0, width=1.\linewidth]{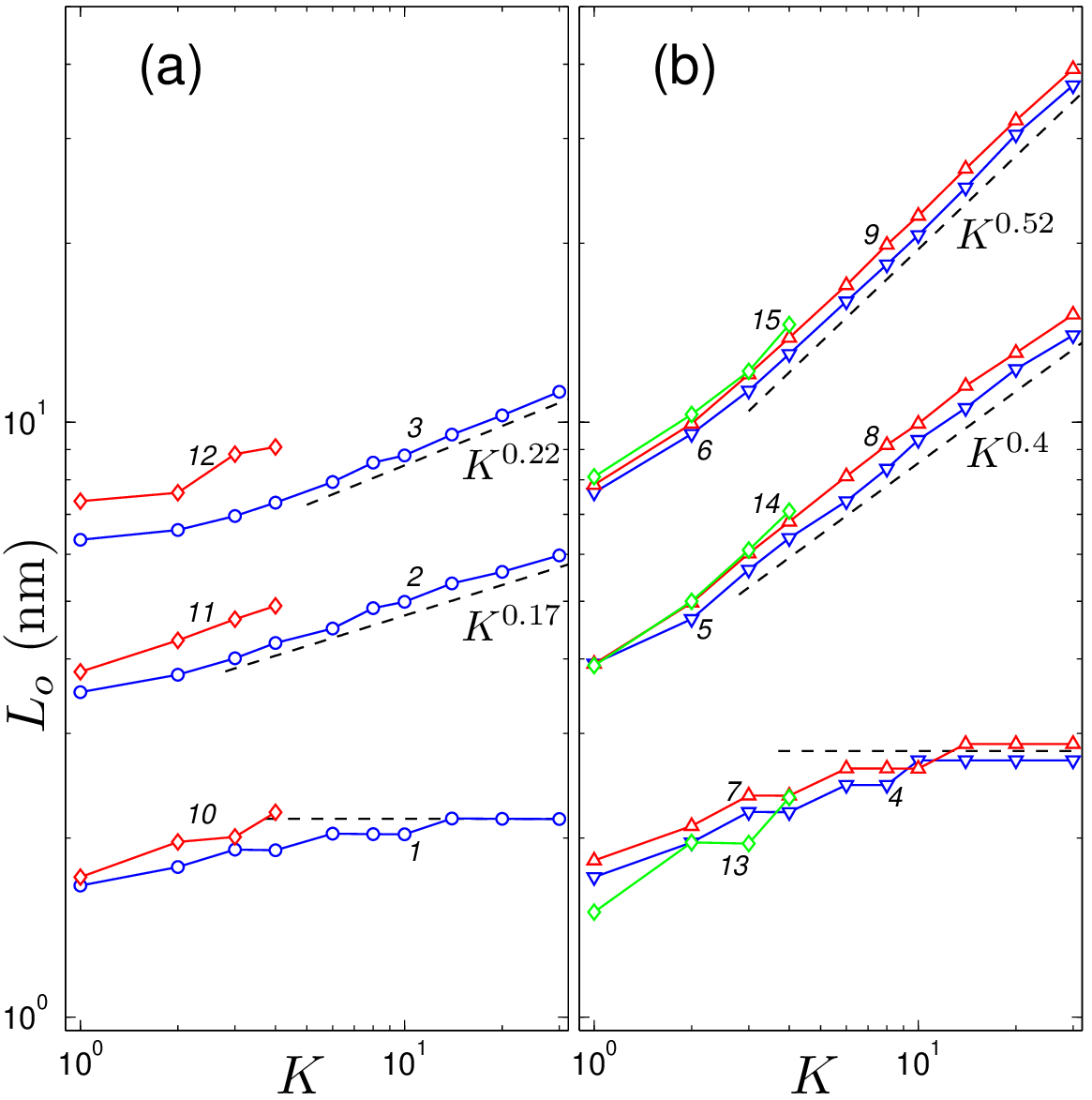}
\end{center}
\caption{\label{fg04}\protect
The dependence of the maximum width of the open state of the slit $L_o$ on the number of layers of the covering sheet $K$.
Within the framework of various models, these dependencies were calculated for a gap with a height of one, two and three layers.
Panel (a) compares the results of the initial two-dimensional model (discreteness parameter $d$=1, curves 1, 2, 3) with the all-atomic LJ-model (curves 10, 11, 12).
Panel (b) shows the values of the maximum width in the two-dimensional model at $d$=2 (curves 4, 5, 6), with $d$=2.129 (curves 7, 8, 9), as well as within the framework of a three-dimensional KC-model (curves 13, 14, 15).
Curves 1, 4, 7, 10, 13 give the dependence for a single-layer gap ($K_g=1$), curves 2, 5, 8, 11, 14 -- for a two-layer gap ($K_g=2$), and 3, 6, 9, 12, 15 -- for a three-layer slit ($K_g=3$).
Logarithmic axes are used, dotted lines show the power-law dependencies of $K^\alpha$.
}
\end{figure}

\section{Critical Slit Width Depending on the Upper Layer Thickness \label{sec3}}

To model a substrate with a linear cavity, consider a system of $K_s=50$ chains with periodic boundary conditions (number of beads in each chain $N_s=200$, 400).
To create a cavity from the top $K_g$ chains, remove $N_g$ beads in its center -- see Fig. \ref{fg03}.
The number of chains $K_g=1$, 2, 3 determines the cavity depth $L_z=K_gh_0$, and the number of beads $N_g$ determines its width $L_x=(N_g+1)a$.
Then the substrate is covered from above with a system of $K$ parallel chains (number of beads $N=160$, 320).
These chains correspond to the $K$-layer graphene sheet covering the cavity in the substrate; free boundary conditions will be used for them.
Thus, with a system of $K_a=K_s+K$ chains consisting of $N_a=K_sN_s-K_gN_g+KN$ beads, we model a linear cavity (slit) of size $L_x\times L_z$ on the flat surface of a graphite crystal, covered from above by a finite multilayer graphene sheet.

To find the stationary state of this molecular system, it is necessary to numerically solve the problem of minimizing its potential energy (\ref{f8}).
To fix the position of the slit, let us fix the $x$-coordinates of the external edge beads of the chains forming the slit.
Solution of the problem (\ref{f8}) showed that for $K_g=1$, 2, 3 there always exists a finite maximum possible cavity width $L_o=aN_g$, at which covering with a graphene sheet does not lead to the sheet filling the cavity space (the slit remains in a stationary open state, where the covering sheet does not touch its bottom) -- see Fig. \ref{fg03} a, c, e.
For greater width (for $L_x>L_o$) the cavity space will always be filled by the covering sheet (covering the cavity leads to the transition of the formed slit to a stationary closed state) -- see Fig. \ref{fg03} b, d, f.
Here, only a stationary state is possible where the covering sheet adheres to the bottom of the slit.
\begin{table}[tb]
\caption{
Dependence of the critical slit width values $L_c$, $L_o$ on the number of layers $K$ in the covering sheet for a two-layer slit ($K_g=2$, $d=2$).
\label{tab2}
}
\begin{tabular}{ccccccccccc}
\hline
\hline
~$K$~      & 1    &  2   &   3  &  4   &   6  &  8   &  10  &  14   &  20   &  30  \\
$L_c$ (nm) & 3.17 & 3.42 & 3.67 & 3.91 & 4.16 & 4.41 & 4.66 &  5.15 & 5.40  &  5.64 \\
$L_o$ (nm) & 3.93 & 4.67 & 5.65 & 6.39 & 7.37 & 8.35 & 9.34 & 10.56 & 12.28 & 14.00 \\
\hline
\hline
\end{tabular}
\end{table}

Note that such a closed state is only possible for slits with width $L_x\ge L_c$.
For narrow slits with $L_x<L_c$, only a stationary open state exists, and for $L_x\in(L_c,L_o)$ there exist two stable stationary states with empty and filled cavity space.
Here, the slit (covered cavity) is a bistable system.
For wide slits with $L_x>L_o$, only a stationary state with filled cavity space exists.
The dependence of the critical slit width values $L_c$ and $L_o$ on the number of layers $K$ in the covering sheet is given in Table \ref{tab2}. 
As can be seen from the table, for a two-layer chain $L_c$, $L_o$ grow monotonically with increasing number of layers $K$.

Figure \ref{fg04} shows the dependence of the maximum width $L_o$ on the number of layers $K$ in the covering sheet.
As can be seen from the figure, for a single-layer slit ($K_g=1$) there is a limitation on the size of the unfilled (open) slit.
Regardless of the number of layers in the covering sheet, the maximum width of the open slit is $L_o\le 2.16$~nm when using the chain model with discreteness $d=1$, $L_o\le 2.70$~nm for $d=2$ and $L_o\le 2.88$~nm for $d=2.129$.
For two- and three-layer slits ($K_g=2$, 3) the maximum width of the open slit increases monotonically with increasing number of layers in the covering sheet as a power function:
$$
L_o\sim K^\alpha,~~\mbox{for}~~K\rightarrow\infty.
$$
For a two-layer slit, the exponent is $\alpha=0.17$, 0.4, and for a three-layer one -- $\alpha=0.22$, 0.52 (values obtained using the chain model with chain discreteness $d=1$ and $d=2, 2.129$).
Thus, increasing the discretization parameter $d$ leads to an increase in the maximum slit width $L_o$.
This width is influenced by the ability of the upper layer to fill the slit, which is related to the possibility of longitudinal movement of the edges of this layer. 
When $d$ increases, this possibility decreases, as the pinning energy increases.

Note that the critical slit width $L_o$ does not depend on the length of the covering chain $L=(N-1)a$, if $N\gg N_g$.
For example, for $K_g=2$, $K=1$, the value $L_o=3.93$~nm ($N_g=15$) is obtained for $N=90$, 180, 360, while for $N=45$ the width is $L_o=3.68$~nm ($N_g=14$), for $N=25$ the width is $L_o=3.44$~nm ($N_g=13$).

The obtained dependencies $L_o(K)$ allow us to conclude that inside a graphite crystal (multilayer graphene sheet), single-layer unfilled linear cavities (empty slits) can have a width of no more than 3~nm, while the width of stable open two-layer and three-layer cavities is practically unlimited.
Note that if the cavity in the substrate was pre-filled with something before covering, for example, with water, then the size of such a cavity is unlimited.
Here, the molecules filling the cavity will prevent its collapse.
\begin{figure}[tb]
\begin{center}
\includegraphics[angle=0, width=0.9\linewidth]{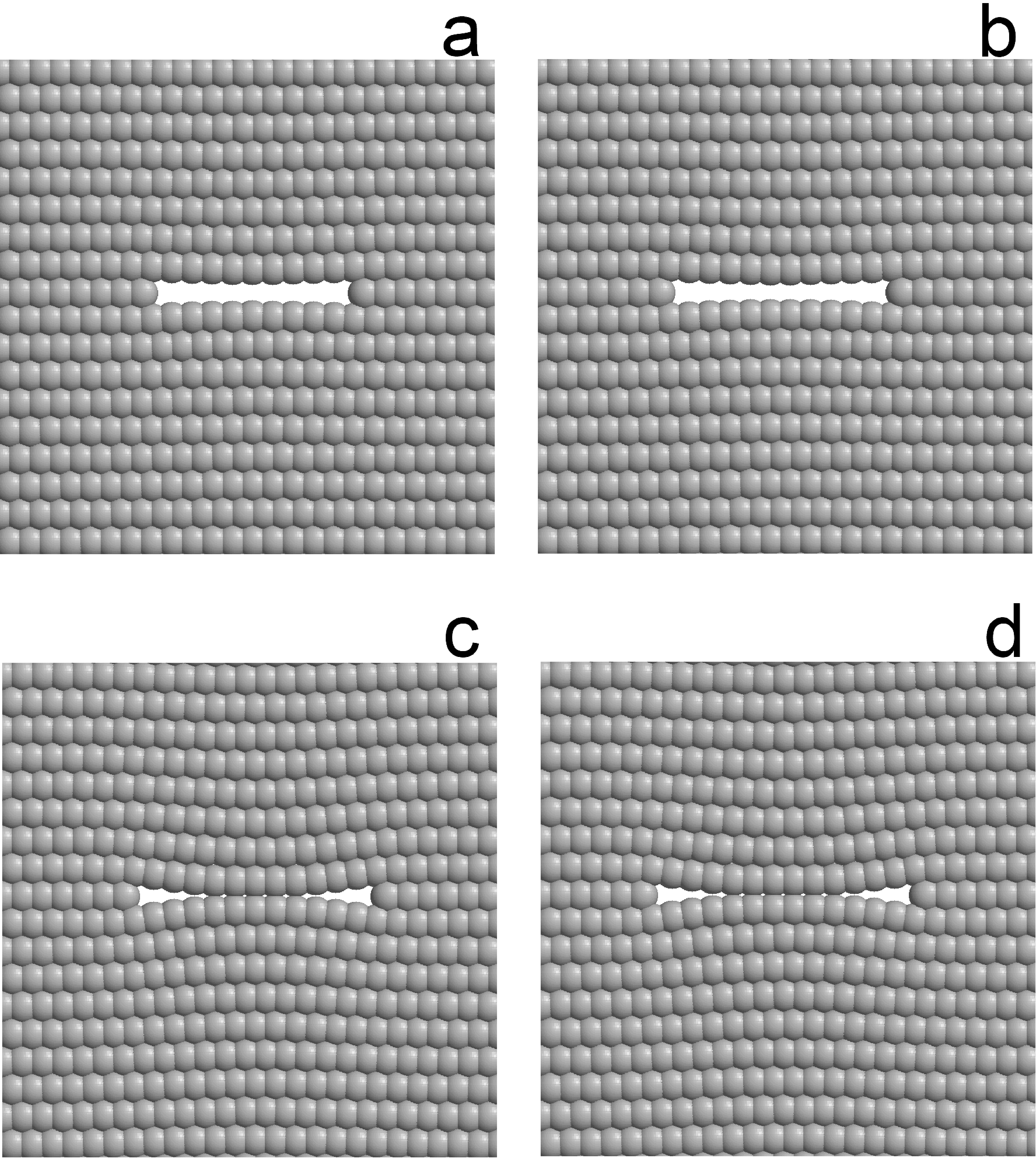}
\end{center}
\caption{\label{fg05}\protect
View of the stationary state of a single-layer slit in a multilayer crystal with width $L_x=aN_g$ for $N_g=10$, 11, 12, 13 (a, b, c, d).
Chain period $a=2a_x$ (chain discreteness $d=2$).
}
\end{figure}

\section{Slits Inside a Multilayer Crystal  \label{sec4}}

To model slits inside a multilayer crystal, consider a system of $K=100$ chains of $N=200$ beads with periodic boundary conditions along both axes.
For definiteness, we will henceforth use the chain model with discreteness $d=2$.

Let us remove $N_g$ beads from $K_g=1$, 2 neighboring chains, thereby creating a slit in the crystal of size $L_x=(N_g+1)a\times L_z=K_gh_0$.
To find the stationary state of the slit, we solve the problem of minimizing the system energy (\ref{f8}).
Numerical solution of the problem showed that a single-layer slit in the crystal can be opened only for $N_g\le 11$ -- see Fig. \ref{fg05}.
Thus, the width of an open single-layer slit is always less than 2.95~nm, which coincides with the estimate obtained in the previous section.
\begin{figure}[tb]
\begin{center}
\includegraphics[angle=0, width=1\linewidth]{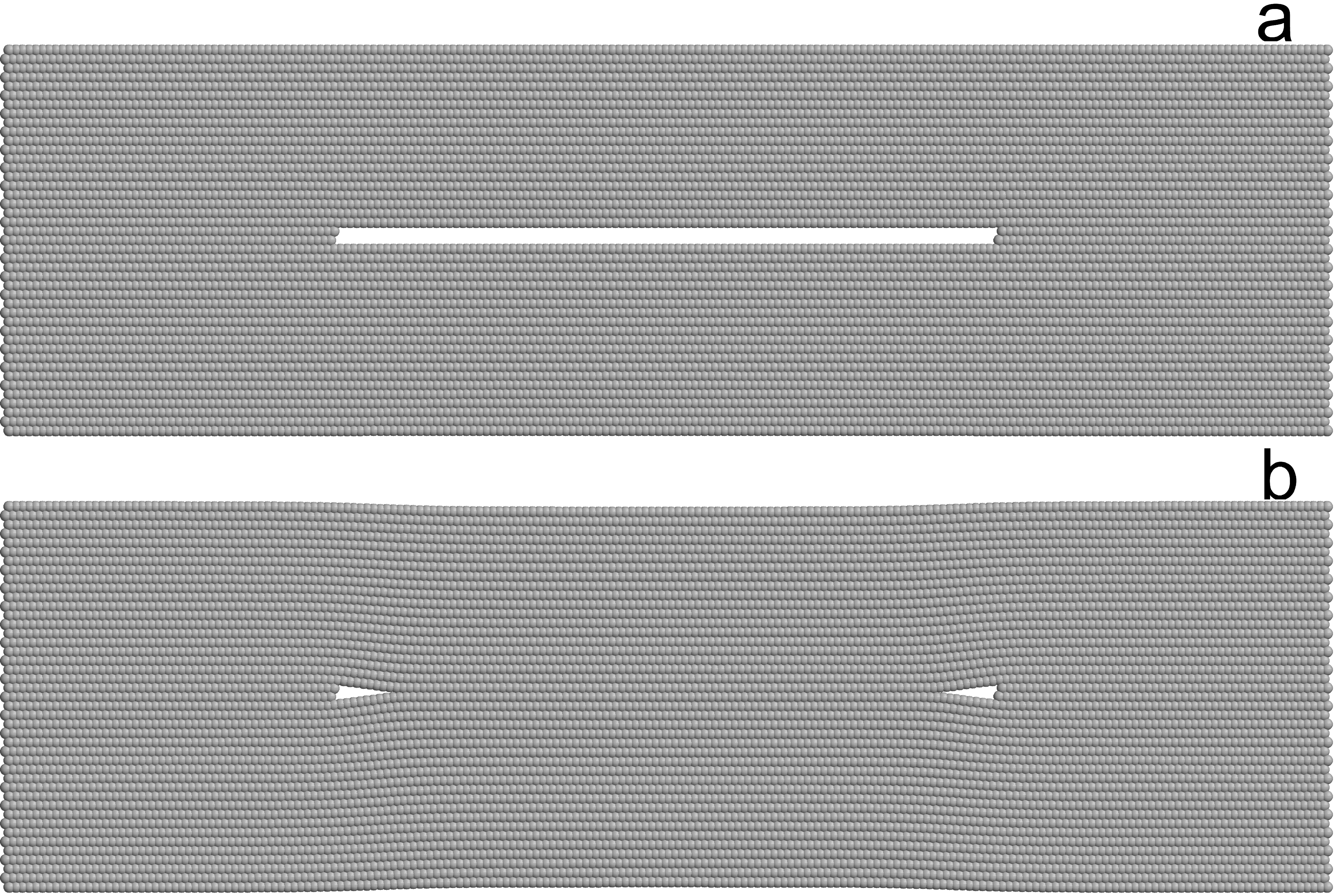}
\end{center}
\caption{\label{fg06}\protect
View of the stationary state of a two-layer (a) open and (b) closed slit in a multilayer crystal with width $L_x=24.8$~nm ($N_g=100$, chain period $a=2a_x$).
}
\end{figure}

\begin{figure}[tb]
\begin{center}
\includegraphics[angle=0, width=1\linewidth]{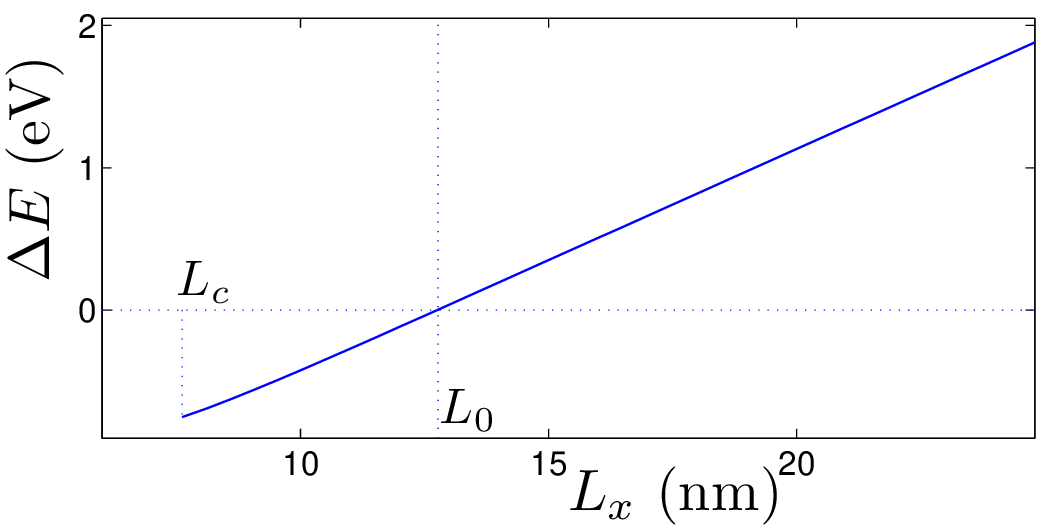}
\end{center}
\caption{\label{fg07}\protect
Dependence of the energy difference $\Delta E=E_o-E_c$ between the open and closed stationary states of a two-layer slit in a multilayer crystal on its width $L_x$.
}
\end{figure}

The solution of problem (\ref{f8}) showed that the open state of a two-layer slit is stable for any length (here the critical value $L_o=\infty$).
For $N_g\ge 30$ (for $L_x\ge L_c=31a=7.61$~nm) there also exists a closed stationary state of the slit, in which its upper and lower surfaces adhere to each other -- see Fig. \ref{fg06}.
The dependence of the energy difference of these stationary slit states $\Delta E=E_c-E_o$ on its width $L_x=(N_g+1)a$ is shown in Fig. \ref{fg07}.
The energy difference increases almost linearly with increasing slit width.
For $L_x<L_0=12.77$~nm the open slit state is energetically more favorable ($E_o<E_c$), while for $L_x>L_0$ the ground state becomes the closed one ($E_o>E_c$).

\section{Influence of Thermal Oscillations \label{sec5}}

To check the stability of the stationary slit states, we have performed molecular dynamics simulation at temperature $T\le 930$~K.
To simulate thermal oscillations of the multilayer structure, we need to numerically integrate the system of Langevin equations
\begin{equation}
M\ddot{\bf u}_{n,k}=-\frac{\partial H}{\partial {\bf u}_{n,k}}-\Gamma M\dot{\bf u}_{n,k}-\Xi_{n,k},~
\label{f9}
\end{equation}
where $k=1,...,K$ denotes the chain index, $n=1,...,N$ denotes the bead index.
Here $M=2M_c$ is the mass of a bead, $\Gamma=1/t_r$ is the friction coefficient (thermostat relaxation time $t_r=10$~ps),
$$
\Xi_{n,k}=(\xi_{n,k,1},\xi_{n,k,2})
$$
is a two-dimensional vector of normally distributed random forces with correlation functions
$$
\langle\xi_{n_1,k_1,i}(t_1)\xi_{n_2,k_2,j}(t_2)=2Mk_BT\Gamma\delta_{k_1k_2}\delta_{n_1n_2}\delta_{ij}\delta(t_2-t_1)
$$
($k_B$ is Boltzmann's constant, $T$ is the thermostat temperature).

In our simulation, we use periodic boundary conditions with a fixed cell size.
Therefore, for correct simulation, it is necessary to account for the thermal expansion of the system along the $z$-axis upon heating when setting the initial cell sizes (compression along the $x$-axis can be neglected).
To study the change in the distance between neighboring graphene layers $\bar{h}(T)$, let us first consider a system of $K=80$ chains of $N=200$ beads.
Take periodic boundary conditions along the $x$-axis with period $Na$, and along the $z$-axis -- free boundary conditions.
As the initial condition for the system of equations of motion (\ref{f9}), take the stationary state of the system of $K$ parallel chains
$$
\{ {\bf u}_{n,k}(0)={\bf u}_{n,k}^0,~~\dot{\bf u}_{n,k}={\bf 0}\}_{n=1,k=1}^{N,~K}.
$$

The system of equations of motion (\ref{f9}) were numerically integrated using the Verlet method \cite{Verle1967} with integration step $\Delta t=1$~fs. 
After equilibrium with the thermostat is reached, the system was simulated for time $t=10$~ns the average distance between neighboring central chains $\bar{h}$ was calculated.

\begin{figure}[tb]
\begin{center}
\includegraphics[angle=0, width=1\linewidth]{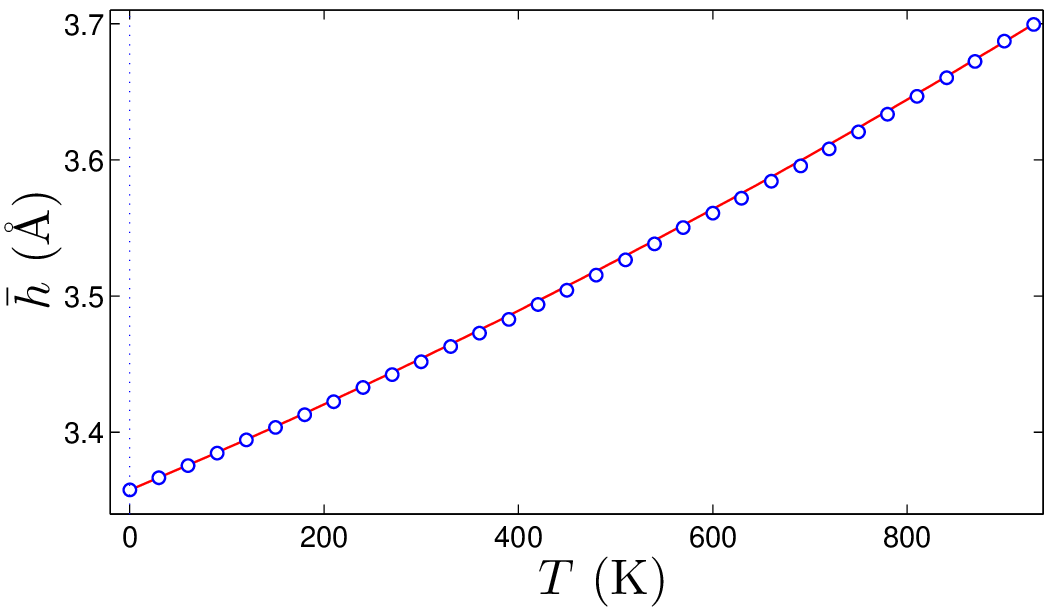}
\end{center}
\caption{\label{fg08}\protect
Dependence of the average distance between neighboring chains $\bar{h}$ on temperature $T$.
Markers show values obtained numerically, the solid line corresponds to dependence (\ref{f10}).
}
\end{figure}

Numerical simulation showed that thermal oscillations lead to an increase in the distance between neighboring chains, described with high accuracy by the formula
\begin{equation}
\bar{h}(T)=h_0+\alpha_1T+\alpha_2T^2,
\label{f10}
\end{equation}
where $h_0=3.3576$~\AA, $\alpha_1=3.0\times 10^{-4}$~\AA/K, $\alpha_2=7.3\times 10^{-8}$~\AA/K$^2$ -- see Fig. \ref{fg08}.

To simulate the influence of thermal oscillations on the slit shape, let us consider a system of $K=100$ parallel chains, located at distance $\bar{h}(T)$ from each other.
Each chain consists of $N=200$, 400, 800 beads.
Along the $x$-axis we used periodic boundary conditions with period $Na$, and along the $z$-axis (by index $k$) -- periodic boundary conditions with period $K\bar{h}(T)$.

Since single-layer slits are unstable when width $L>3$~nm, we will consider next more interesing case of slits with bigger height.

To create a two-layer slit of width $L=(N_g+1)a$, let us remove $N_g$ beads from two neighboring chains.

\begin{figure}[tb]
\begin{center}
\includegraphics[angle=0, width=1.0\linewidth]{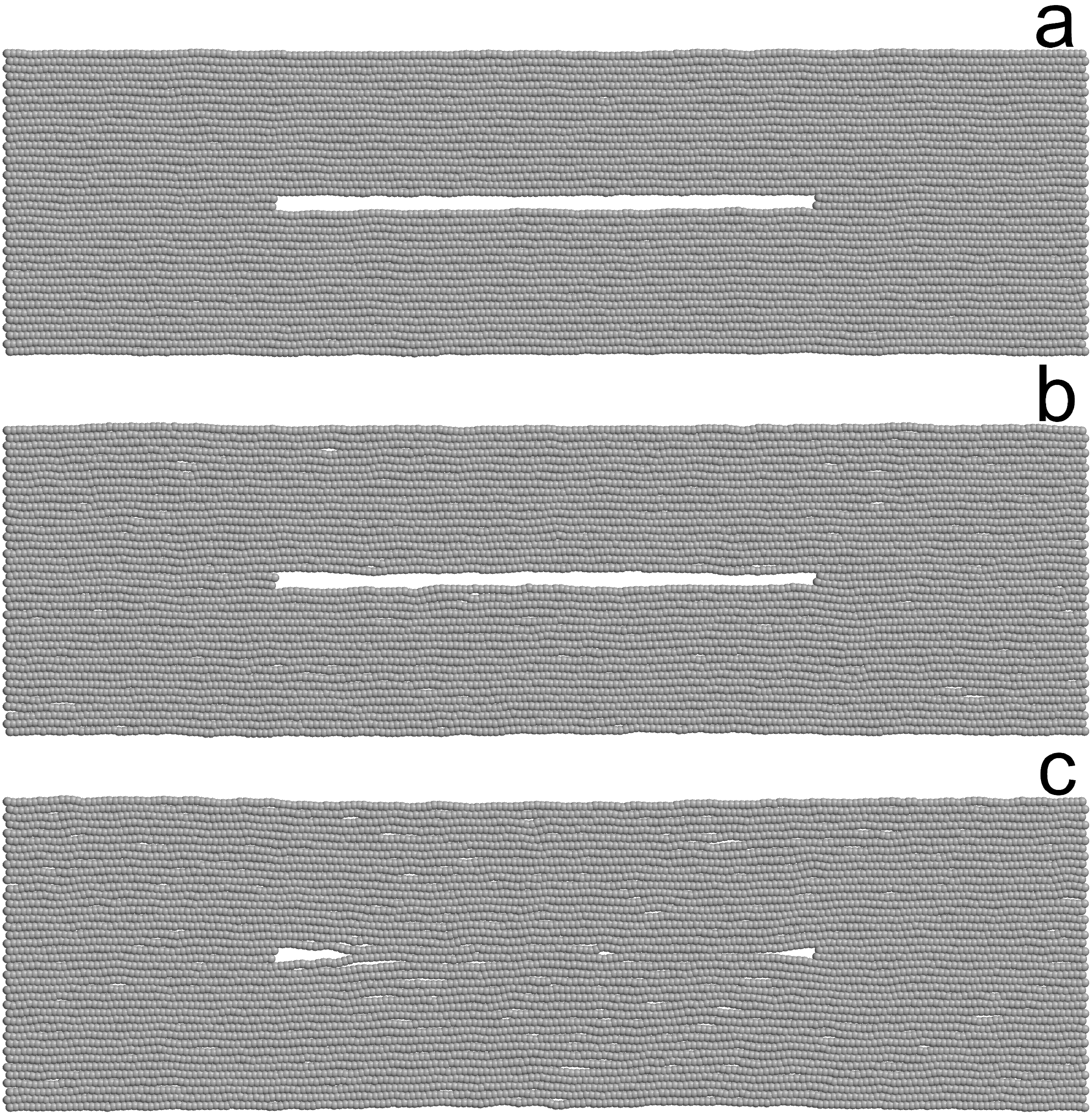}
\end{center}
\caption{\label{fg09}\protect
State of a two-layer slit ($K_g=2$, $N_g=100$) in a multilayer crystal at temperature (a) $T=300$, (b) 510 and (c) 540~K.
The configuration of the multilayer system in the vicinity of the slit at time $t=5$~ns is shown.
}
\end{figure}

Numerical integration of the system of equations of motion (\ref{f9}) showed that a slit with width $L< L_0=15$~nm ($N_g<60$) always remains in the open state for all temperature values.
Wider slits at high temperatures will transition to the closed state.
For example, for $N_g=50$ the planar slit remained open at $T\le 930$~K during the entire simulation time $t=5$~ns.
For $N_g=60$ the slit remains open at $T\le 600$~K (at higher temperatures the slit periodically transitions from the open state to the closed state and back).
For $N_g=70$, 80, 100 and 200 ($L=17.4$, 19.9, 24.8 and 49.4~nm) the slit remains open at temperature $T\le T_c=510$, 630, 510 and 420~K, respectively.
Here, for $T>T_c$ the slit transitions to the closed state and thereafter remains in it -- see Fig. \ref{fg09}.

A three-layer slit ($K_g=3$) is always stable against thermal oscillations.
For example, at any temperature $T\le 930$~K it always remains in the open state.

\section{All-Atom Simulation \label{sec6}}

It is natural to compare the predictions of the two-dimensional graphene model with a more detailed three-dimensional atomic model, where an individual particle corresponds to a carbon atom, not a strip of atoms of width $da$.
We assume that hydrogen atoms are attached to the carbon atoms at the free edges of the graphene sheet.
This CH group will be treated as one (unified) atom of mass $13m_p$.

The potential energy of a $K$-layer graphene nanoribbon on a flat substrate has the form:
\begin{eqnarray}
E=\sum_{k=1}^K\sum_{n=1}^N[E_{n,k}+P(z_{n,k})]\nonumber\\
+\sum_{k_1=1}^{K-1}\sum_{k_2=k_1+1}^K\sum_{n_1=1}^N\sum_{n_2=1}^N W(r_{n_1,k_1;n_2,k_2}),
\label{f11}
\end{eqnarray}
where the vector ${\bf u}_{n,k}=(x_{n,k},y_{n,k},z_{n,k})$ specifies the coordinates of the $n$-th carbon atom of the $k$-th nanoribbon, $N$ is the number of atoms in each layer.

The first term in the sum (\ref{f11}) $E_{n,k}$ gives the interaction energy of the $n$-th atom of the $k$-th nanoribbon with neighboring atoms of the nanoribbon (accounting for deformations of valence bonds, valence and torsional angles \cite{Savin2010prb}).
The potential
\begin{equation}
P(z)=e_0[\beta (h_z/z)^\alpha-\alpha(h_z/z)^\beta]/(\alpha-\beta),
\label{f12}
\end{equation}
describes the interaction energy of a nanoribbon atom with the flat substrate $z\le 0$, formed by the surface of a graphite crystal, interaction energy $e_0=0.0518$~eV, equilibrium distance to the substrate plane $h_z=3.37$~\AA, exponents $\alpha=10$, $\beta=3.75$ \cite{Savin2019ftt}.
\begin{figure}[tb]
\begin{center}
\includegraphics[angle=0, width=1.0\linewidth]{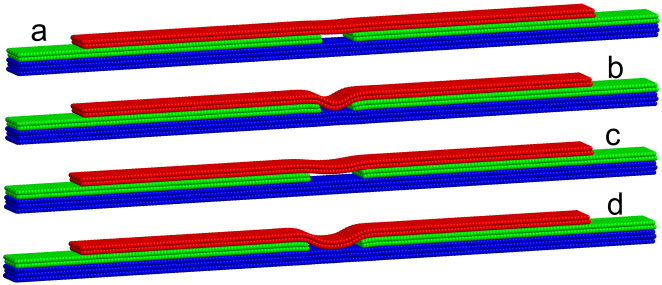}
\end{center}
\caption{\label{fg10}\protect
Stationary state of a two-layer slit ($K_g=2$), covered by a two-layer graphene sheet for width $L=2.70$~nm ($N_g=21$): (a) open and (b) closed state; for width $L=4.30$~nm ($N_g=34$): (c) open and (d) closed state.
The upper graphene sheet is shown in red, the substrate sheets participating in the slit formation are shown in green.
The LJ model is used.
}
\end{figure}

The last term in formula (\ref{f11}) describes the energy of non-valent interaction between atoms of different layers, $r_{n_1,k_1;n_2,k_2}=|{\bf u}_{n_2,k_2}-{\bf u}_{n_1,k_1}|$ is the distance between atoms $n_1,k_1$ and $n_2,k_2$, potential
\begin{equation}
W(r)=\epsilon_w\{[(r_w/r)^6-1]^2-1\},
\label{f13}
\end{equation}
where $\epsilon_w=0.002757$~eV, $r_w=3.807$~\AA~ \cite{Setton1996}.

As discussed in Section \ref{sec2}, the model described above with pairwise Lennard-Jones interactions of atoms (LJ-model) underestimates pinning energy.
Therefore, let us also consider a force field (KC-model), in which the pinning energy of graphene sheets agrees with the results from DFT \cite{Ouyang2018nl}.
In this force field, the interaction of nearby atoms is determined by the REBO potential \cite{Brenner2002}, and non-valent interactions of atoms from different layers are described by the Kolmogorov-Crespi potential \cite{Kolmogorov2005}.
The potential energy of the system was supplemented by a term accounting for interaction with the substrate (\ref{f12}).
Energy minimization calculations for this model were performed in the LAMMPS package \cite{Plimpton1995}.

Let us take as the multilayer substrate a nanoribbon of $K_s=5$ layers, where the "zigzag" direction coincides with the $x$-axis, and the "armchair" direction coincides with the $y$-axis.
Consider nanoribbons of size $L_x=2N_xa_x\times L_y=3N_yr_c$, where $N_x=200$, $N_y=5$. Each layer of such a nanoribbon will consist of $N_s=4N_xN_y=4000$ carbon atoms. 
When modeling the substrate, we used periodic boundary conditions with periods along the $x$ and $y$ axes of $2N_xa_x=49.12$~nm and $3N_yr_c=2.127$~nm.

\begin{figure}[tb]
\begin{center}
\includegraphics[angle=0, width=0.96\linewidth]{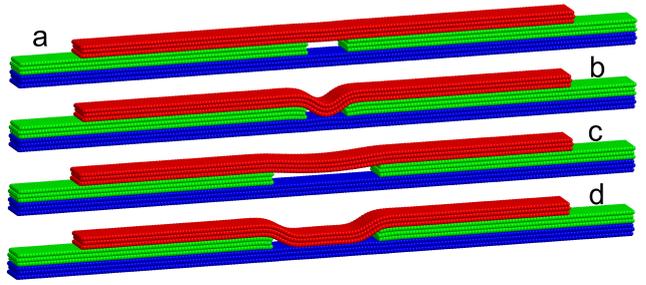}
\end{center}
\caption{\label{fg11}\protect
Stationary state of a three-layer slit ($K_g=3$), covered by a three-layer graphene sheet for width $L=3.44$~nm ($N_g=27$): (a) open and (b) closed state; for width $L=8.72$~nm ($N_g=70$): (c) open and (d) closed state.
The upper graphene sheet is shown in red, the substrate sheets participating in the slit formation are shown in green.
The LJ model is used.
}
\end{figure}

To create a slit from the top $K_g=1$, 2, 3 nanoribbons, let us remove $N_g\times 4N_y$ atoms in their center, thereby creating a transverse cavity of width $L=(N_g+1)a_x$ -- see Figs. \ref{fg10} and \ref{fg11}.
Then we covered the substrate with the cavity with a $K$-layer nanoribbon of the same width $L_y$, but shorter length $(2N-1)a_x$ with the number of transverse unit cells $N=160$.
Here, we used free boundary conditions along the $x$-axis, and periodic boundary conditions along the $y$-axis.
This three-dimensional molecular structure will correspond to the two-dimensional multilayer structure considered in Section \ref{sec3}.

To find the stationary state of the slit covered by a multilayer graphene sheet, one must numerically solve the problem of minimizing the potential energy
\begin{equation}
E\rightarrow\min: \{ {\bf u}_{n,k}\}_{n=1,k=1}^{N_{k},K_a},
\label{f14}
\end{equation}
where the total number of layers is $K_a=K_s+K$, $N_k$ is the number of atoms in the $k$-th layer (nanoribbon).
\begin{table}[b]
\caption{
Dependence of the critical slit width values $L_c$, $L_0$, $L_o$ on the number of layers $K$ in the covering sheet for a $K_g$-layer slit (values are given in nm, the LJ-model is used).
\label{tab3}
}
\begin{tabular}{c|ccccc}
\hline
\hline
~$K_g$~ & ~$K$~  & 1    &  2   &   3  &  4   \\ \hline
   1    & $L_o$  & 1.72 & 1.97 & 2.09 & 2.21 \\
   2    & $L_c$  & 2.09 & 2.46 & 2.58 & 2.82 \\
   2    & $L_0$  & 2.58 & 2.95 & 3.19 & 3.44 \\
   2    & $L_o$  & 3.81 & 4.30 & 4.67 & 4.91 \\
   3    & $L_c$  & 2.09 & 2.46 & 2.58 & 2.82 \\
   3    & $L_0$  & 3.56 & 4.05 & 4.42 & 4.79 \\
   3    & $L_o$  & 7.37 & 7.61 & 8.37 & 9.09 \\
\hline
\hline
\end{tabular}
\end{table}

Numerical solution of problem (\ref{f14}) showed that for the number of layers forming the slit $K_g=1$, 2, 3, there always exists a maximum possible slit width $L_o$, for which a stable stationary state with an open slit exists (in this state the covering sheet does not touch the slit bottom) -- see Fig.~\ref{fg10} a, c and \ref{fg11} a, c.
For greater width, only a closed stationary state of the slit is possible, where the covering sheet closely adheres to the slit bottom -- see Fig.~\ref{fg10} b, d and \ref{fg11} b, d.
The closed stationary state exists only for slits with width $L>L_c$, where the critical slit value is $L_c<L_o$.
Thus, for width $L<L_c$ only the open stationary state of the slit exists, for $L\in( L_c,L_o)$ two stationary states exist simultaneously (open and closed), and for $L>L_o$ only the closed state exists for the slit.
The open stationary state is energetically more favorable only for slit width $L<L_0$, where the value $L_0\in(L_c,L_o)$. 
Characteristic slit width values $L_c$, $L_0$ and $L_o$ are given in Table~\ref{tab3}.

As can be seen from Figure \ref{fg04}, the three-dimensional LJ-model agrees well with the two-dimensional chain model ($d=1$), and the KC-model agrees with the same model for $d=2.219$.
The values of the maximum slit width in the KC-model, as expected, slightly exceed the estimates from the LJ-model.
Differences in the pinning energy of the two three-dimensional models also affect the deformation changes of the upper graphene sheet upon slit collapse.
For example, in the KC-model with stronger pinning, the collapse of a single-layer slit leads to a shift of the edges of the upper layer relative to the middle layers by 0.1~\AA~ compared to 1~\AA~ for the LJ-model.
Similar, but less pronounced differences are observed during the collapse of two-layer (shift of 0.4~\AA~ vs. 1.3~\AA) and three-layer slits (shift of 1.5~\AA~ vs. 3.5~\AA).

\section{Conclusion  \label{sec7}}
In this work, the previously proposed two-dimensional chain model of multilayer graphene sheets \cite{Savin2023pe} was refined.
By increasing the chain period by a factor of $d=2.129$ and rescaling the interaction constants, the mechanical properties and cohesion energy of graphene sheets were preserved, and the pinning energy was reconciled with DFT values \cite{Ouyang2018nl}.

Using the two-dimensional coarse-grained chain model, the formation of planar slits in multilayer graphite crystals was simulated.
It was shown that when covering a linear cavity on the flat surface of a graphite crystal with a multilayer graphene sheet, an open (unfilled slit) can form only if the cavity width does not exceed a critical value $L_o$. 
For greater cavity width (for $L>L_o$), the graphene sheet, due to bending and edge shifting, completely adheres to its bottom, forming a closed (collapsed) slit state. 
A closed state is only possible for slit widths exceeding a threshold value $L_c<L_o$.
Thus, for narrow slits with width $L<L_c$, only an open stationary state is possible.
For slits of medium size $L\in(L_c,L_o)$, two stable stationary states exist -- open and closed.
Here, the slit (covered cavity) is a bistable system.
And for wide slits with $L>L_o$, only a closed stationary state can exist, with the cavity space filled by the covering sheet.
The critical width of the open slit $L_o$ increases monotonically with the number of layers $K$ in the covering sheet.
For a single-layer cavity, there is a finite critical value of its width $L_o<3$~nm, independent of $K$.
For two- and three-layer cavities, the maximum width of the open slit increases with increasing $K$ as a power function: $L_0\sim K^\alpha$, as $K\nearrow\infty$ (exponent $0<\alpha<1$).

The conducted simulation showed that a single-layer slit inside a graphite crystal can be in an open (unfilled) state only if its width is $L<3$~nm.
Two- and three-layer slits here can have stable open states at any width (critical value $L_o=\infty$).
For $L>7.6$~nm, a two-layer slit can also be in a stationary closed state, where its lower and upper surfaces adhere to each other.
The energy difference of these stationary states increases linearly with increasing slit width.
For width $L<L_0=12.8$~nm the open state has lower energy, and for $L>L_0$ -- higher energy.

Simulation of thermal oscillations showed that open states of two-layer slits with width $L<15$~nm
are stable against thermal oscillations at temperature $T\le 930$~K.
Wider slits at $T>400$~K transition from the open to the more energetically favorable closed state. Open states of three-layer slits are always stable against thermal oscillations.

The simulation of slits using three-dimensional all-atom models confirmed the main results obtained using the two-dimensional chain model.
\\ \\
{\bf Funding}

The study was supported by the Russian Science Foundation grant No. 25-73-20038,
\url{https://rscf.ru/project/25-73-20038/}.
\\ \\
{\bf Conflict of interest}\\

The authors declare that they have no conflict of interest.
\\

\end{document}